\documentclass{article}
\usepackage{graphicx}
\usepackage{stywhispers,amsmath,epsfig}

\title{detection and identification accuracy of PCA-accelerated real-time processing of hyperspectral imagery}
\name{Abigail Basener and Meagan Herald}
\address{Department of Applied Mathematics, Virginia Military Institute, Lexington, Virginia 24450 USA }
\begin{document}
%
\maketitle
\begin{abstract}
Real-time or near real-time hyperspectral detection and identification are extremely useful and needed in many fields. These data sets can be quite large, and the algorithms can require numerous computations that slow the process down. A common way of speeding up the process is to use principal component analysis (PCA) for dimension reduction. In the reduced dimensional space, provided by a subset of the principal components, fewer computations are needed to process the data resulting in a faster run time. In this paper, we propose a way to further decrease the time required to use PCA by investigating how many principal components may be omitted with minimal impact on the detection rate.  Using ACE to perform the detection, and then probability, and spectral fit for identification, we find that the number of principal components can be reduced by a substantial amount before seeing a noticeable change in detection rates. 
\end{abstract}
\section{Introduction}
\label{sec:intro}

Military, industry, and humanitarian organizations use hyperspectral imagery to detect and identify targets of interest \cite{loughlin2020efficient}. Ideally, the identification occurs in real-time; however, the necessary data cleaning and detection algorithms can be both time and resource costly making the real-time aspect slower than desired. While there are different ways to adjust the algorithms and data to speed up calculations, this usually is accomplished by reducing either the number of computations or the size of the data. Dimension reduction is a method of projecting a data set into a lower dimensional space without losing valuable information.  One common way of doing dimension reduction is principal component analysis (PCA)\cite{basener2018}. 

To help speed up computations we want to use a minimal number of principal components (PCs) without sacrificing information.  Standard PCA algorithms naturally ordered the PCs by importance and thus each additional PC used has diminishing returns.  We propose that the latter PCs may not be worth the time to process them. The question then becomes, when to stop calculating PCs. \cite{osti_1638363}. In this paper, we will look at the target detection rates and identification accuracy when we change only the number of PCs. 

\section{Data Summary}
\label{sec:format}

For this research, the hyperspectral data collected from Aberdeen Proving Ground's H-Field from August 18th to 31st 1995 was used. This data set has a total of 77 different types of targets, which are a mix of vehicles and target panels. The data was collected with the Hyperspectral Digital Imagery Collection Experiment (HYDICE). It collected 210 channels from 0.4 $\mu m$ to 2.5 $\mu m$ \cite{Olsen1997}. After removing bands that were hindered by the atmosphere there were 145 bands, each of which creates a principal component. 

\section{Detection Summary}
\label{sec:pagestyle}

ACE, a well-established algorithm for target detection, stands for adaptive coherence estimator and is defined by Formula \ref{eq:Ace} where $\sum$ is the covariance matrix, $t$ is the target of interest to detect, and $x$ is the pixel you are looking at. \cite{basener2011automated}

\begin{equation}
ACE(x) = \frac
{x\sum{t}^{-1}}
{\sqrt{x\sum{x}^{-1}}\sqrt{t\sum{t}^{-1}}}
\label{eq:Ace}
\end{equation}

This formula returns a score of how likely the pixel contains the target material. If the ACE score passes our threshold, set to 0.5, then we labeled that position a region of interest (ROI). Once a ROI has been detected as a possible target it is passed to an identification step where spectral fit and probability techniques are used\cite{basener2017ensemble} \cite{basener2011automated}. In this identification step, the pixel is then compared against the background and confusers as well as the target spectra. The background is found by looking at pixels around the ROI with low ACE scores and confusers are a set of objects known to look similar to the target.  This step concludes with the pixel classified as either a real target or a non-target.

It is important to remember that this sample is strictly looking at man-made materials therefore it may not transfer well to other types of targets such as organic materials. However, with 77 different types of targets, we believe that this data set provides a good sample of varying object materials and is representative of most hyperspectral datasets.

\section{PCA summary}
\label{sec:typestyle}

Principal component analysis determines which ratio of variables creates the most variance in the data set and therefore the most important direction of spread, which is designated as the first PC. Then the algorithm looks for the next largest variance by looking at the directions perpendicular to all previous PCs.  Once the next PC is determined, the process continues. 

The PCA, shown in Formulas \ref{eq:Wm} and \ref{eq:Am}, is used to find the column vector which is called the principal component of the image. There is a column for each PC that is used.  
\begin{equation}
W = P \sqrt{D^{-1}}
\label{eq:Wm}
\end{equation}
The diagonal matrix, $D$, has the eigenvalues of the covariance matrix on the diagonal. The $P$ matrix is built such that its column vectors are the eigenvectors of the covariance matrix. We can use the $W$ matrix for the whitening transform shown in Formula 3. This also requires the average of the image’s spectra, $\mu$, and $a$ the particular spectra of interest. \cite{basener2018}

\begin{equation}
\hat{a} = W^{T}(\textbf{a}-\mu)
\label{eq:Am}
\end{equation}

Only considering directions perpendicular to the previous PCs, the algorithm ensures any new PCs do not include variance in the previously determined directions. This implies the current PC focuses on variance in unaccounted for directions by the PCs which preceded it.

\section{Primary Results}
\label{sec:majhead}

\begin{figure}[htp]
\includegraphics[width=0.5\textwidth]{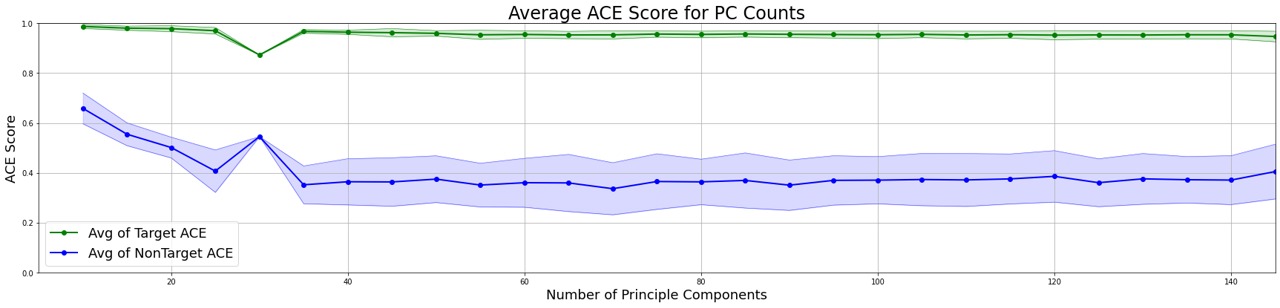}
\caption{A envelope plot of ACE scores for targets and non-targets across PC counts ranging from 5 to 145 increasing by increments of 5.}
\label{fig:AvgEn}
\end{figure}

\begin{figure}[htp]
\includegraphics[width=0.5\textwidth]{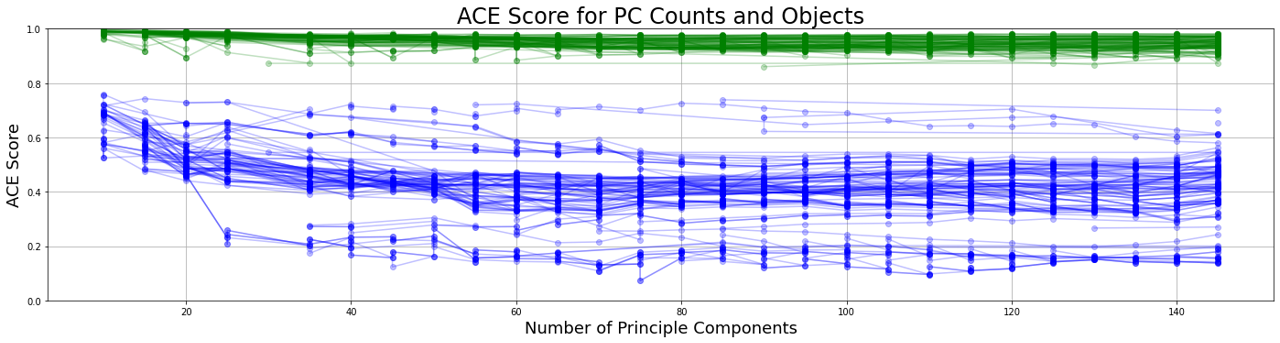}
\caption{The ACE scores for targets and non-targets across PC counts ranging from 5 to 145 by increments of 5 for each individual target.}
\label{fig:AvgAll}
\end{figure}

The PCA already orders components by importance and the principal components which are calculated near the end of the process may be omitted to provide a more efficient run time with the goal of providing real-time identification. We ran the ACE algorithm to perform the detection on the data while varying the PC numbers from 5 to 145 in increments of 5 PCs. In the Figures \ref{fig:AvgEn} and \ref{fig:AvgAll}, the ACE scores are shown for each number of PCs used. In Figure \ref{fig:AvgEn} the green line represents the mean of the ACE score of a real target found, with the shaded region being one standard deviation away from the mean.  In this case, the ROIs were confirmed by the identification step. The blue line is the mean non-target ACE score, plotted with a shaded region of one standard deviation in either direction.   With the blue line, the ROIs were determined not to be real targets and were false detections.  

There is an inconstancy seen with the usage of 30 PCs. In this case, the ACE and identification methods only found a single ROI. Our initial analysis showed this was a result of how the algorithms worked with this data set and the drop with subsequent jump should be ignored for the purposes of this paper. 

To look at this analysis in more detail, the exact ACE score for a single object is shown in Figure \ref{fig:AvgAll} while the number of PCs used varies. Each green line represents a different target, and each blue line represents a different non-target. 

Overall, there is a considerable difference between 5 and 10 PCs looking over a 5 PC increment; however, there is a minimal difference between 100 and 105 PCs over the same sized increment. This validates our earlier claim that we do not need to calculate all the PCs to detect and identify the ROI.  Nevertheless, there does seem to be some optimal number of PCs, in this case around 40 PCs, which are required for accurate classification. Also, note that the number of PCs does not affect the target ACE score as much as it impacts the ACE score for non-targets.  In cases where the hyperspectral detection is strictly used to detect targets, a lower number of PCs may be used.   

\subsection{Visual Explanation }
\label{ssec:subhead}

\begin{figure}[htp]
\includegraphics[width=0.5\textwidth]{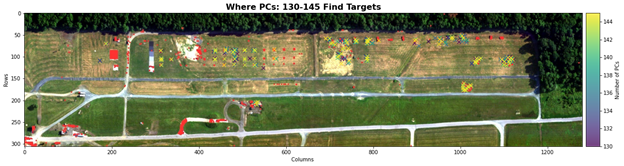}
\caption{ROIs plotted on the ground truth for PC counts from 130 to 145 compared with the full 145 PCs.}
\label{fig:ROI130}
\end{figure}

\begin{figure}[htp]
\includegraphics[width=0.5\textwidth]{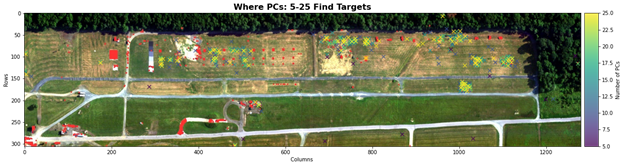}
\caption{ROIs plotted on the ground truth for PC counts from 5 to 25 compared with the full 145 PCs.}
\label{fig:ROI5}
\end{figure}

To further observe the differences between the PCs used and targets or non-targets identified we plotted the ROI locations on the visible image in Figures \ref{fig:ROI130} and \ref{fig:ROI5}.  The red pixels in the image are all the locations with an ACE score above our threshold of 0.5 given the full number of PCs.  Recall, not all of these ROIs will pass the identification step, which is why objects like parts of the road are detected even though there is not an actual target there.  Each $\times$ in the image represents the center of a ROI and the color denotes the number of PCs used to find the ROIs.  Ideally, the ROIs should be on the red marks. When an ROI is found on a part of the image that is not on top of a red pixel, the process is finding a false detection which would not have found using the full 145 PCs. In general, when the PC count is too low, the false detection rate rises. 

To process the data in a reasonable amount of time the number of ROIs processed for identification were limited to the first 100. This is the reason there are targets that went undetected. The scope of this paper was to compare how the number of PCs used affects the ROIs and not to identify every target, using a 100 ROIs limit will not affect the outcome.

Figure \ref{fig:ROI130} shows  the results when using 130 to 145 PCs. In this case, the ROIs tend to be lined up on the targets most notably in the center of the figure where the targets are positioned in rows and columns.  Also, note there is not a large difference between the purple, green, and yellow marks, indicating that in the 130 to 145 PC range, including the 15 additional PCs did not improve the detection of ROIs by much.

We can then look at the same view, using 5 to 25 PCs in Figure \ref{fig:ROI5}. We can see the ROIs are much more scattered correlating with our previous results, Figure \ref{fig:AvgEn}. In Figure \ref{fig:ROI5}, the purple, green, and yellow marks are more spread out implying, the increment from 5 to 25 PCs is important and cannot be ignored. In addition, the false detection rate is much higher.  

There is a significant difference in the ACE scores for a lower number of PCs; however, at some point, increasing the number of PCs does not affect the output of the detection step. This is redundancy that should be removed to minimize the computations needed for detection. 
 
\subsection{Individual Targets}
\label{sssec:subsubhead}

\begin{figure}[htp]
\includegraphics[width=0.5\textwidth]{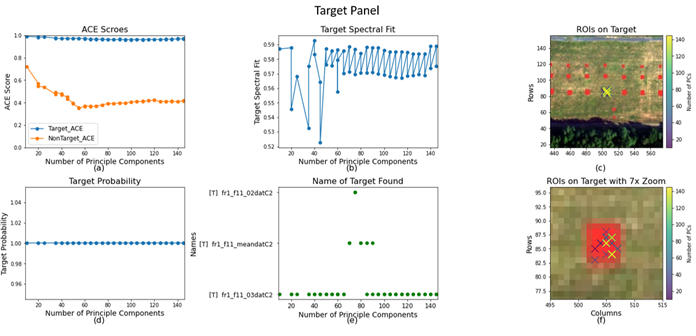}
\caption{An example of a standard target panel for a single target to demonstrate how this target's detection and identification change for different PC used.}
\label{fig:tarNorm}
\end{figure}

\begin{figure}[htp]
\includegraphics[width=0.5\textwidth]{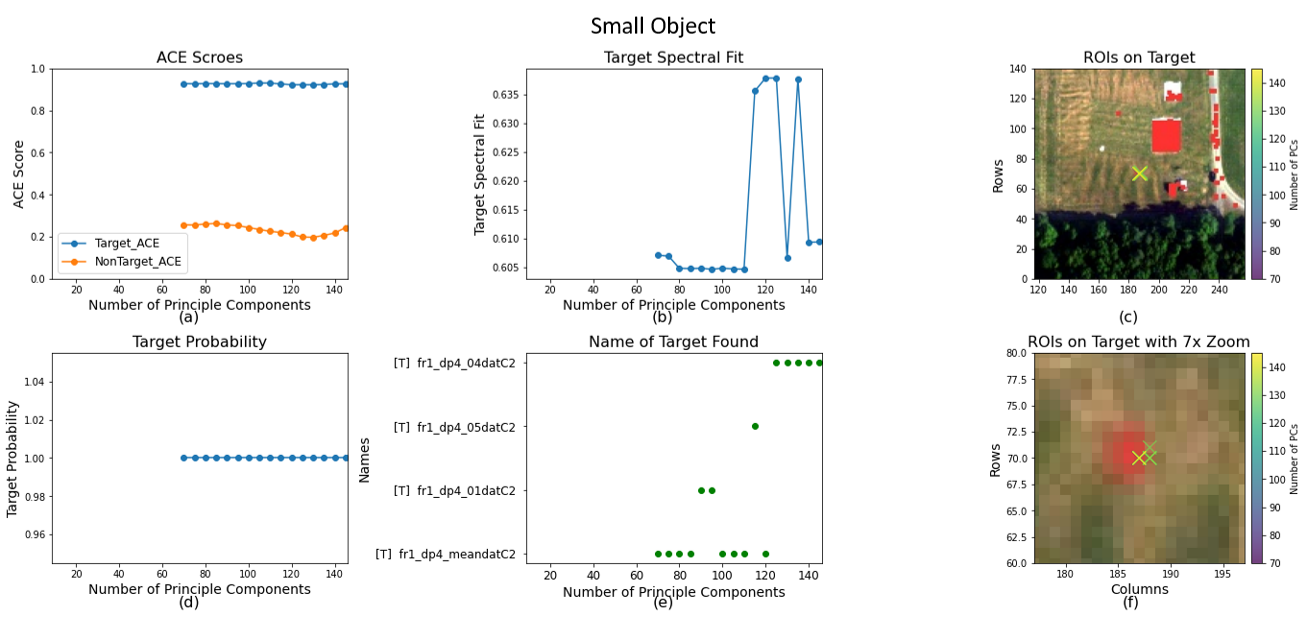}
\caption{A small object completely undetected when fewer than 70 PCs are used.}
\label{fig:tarSP}
\end{figure}

\begin{figure}[htp]
\vspace{-5pt}
\includegraphics[width=0.5\textwidth]{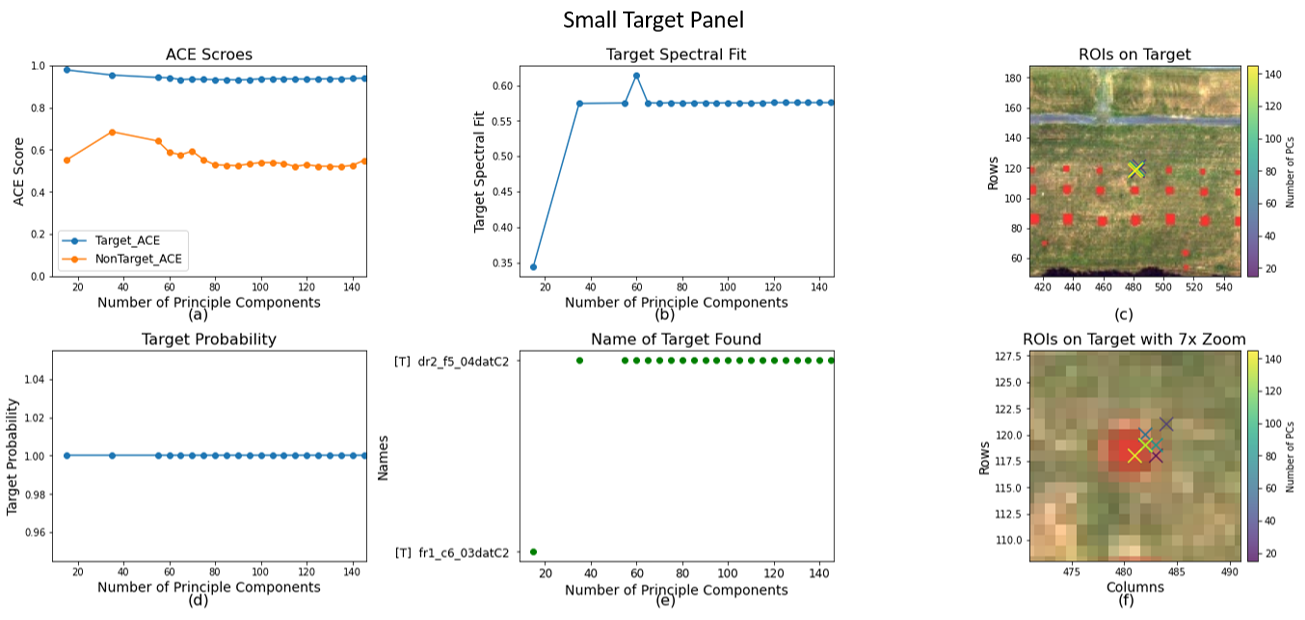}
\caption{A small target with good detection and identification in higher PC levels which becomes harder to detect and identify below 40 PCs.}
\label{fig:tarSG}
\end{figure}

\begin{figure}[htp]
\includegraphics[width=0.5\textwidth]{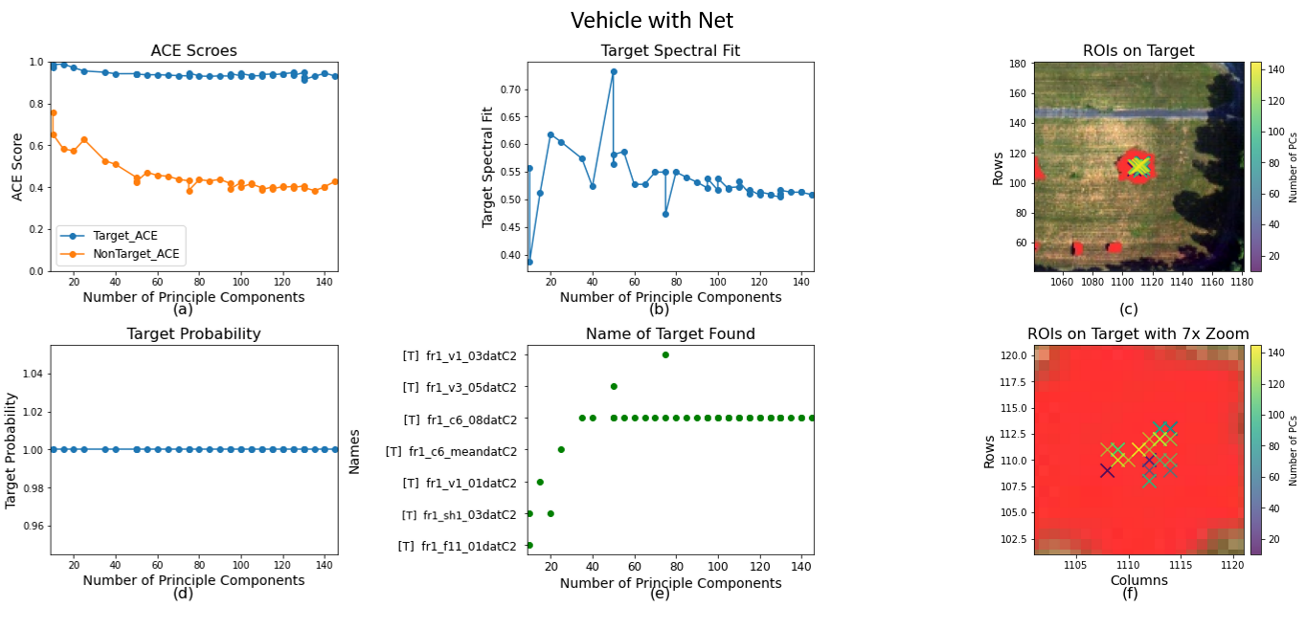}
\caption{A vehicle target with two materials demonstrates identification issues below 40 PCs.}
\label{fig:tarTanCam}
\end{figure}

\begin{figure}[htp]
\includegraphics[width=0.5\textwidth]{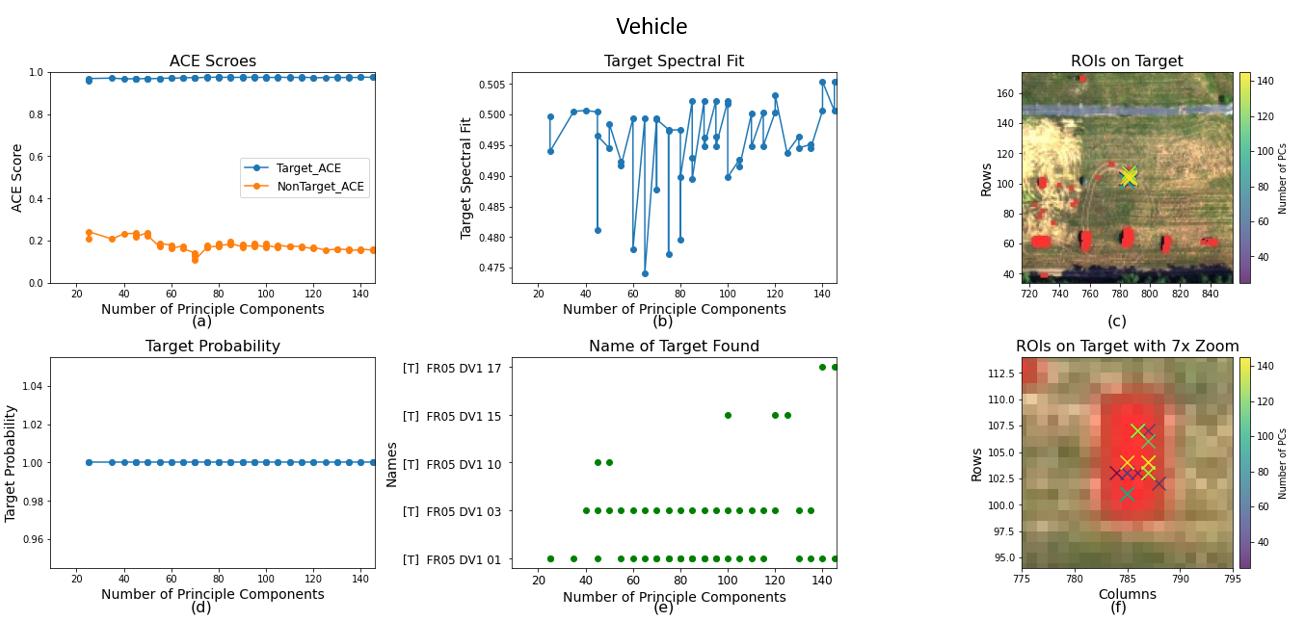}
\caption{A vehicle target with good detection though poor identification for all levels of PCs used. }
\label{fig:tarveh}
\end{figure}

Lastly, we looked how particular targets are affected by the number of PCs used. We took each ROI that was identified as a target when using the full 145 PCs and called each one an object. Then we looked at every ROI found within $\pm$ 3 pixels of the center of the object for any PC amount. We plotted an $\times$ in the center of each of these ROIs on the following figures. Since we are only considering a single object for each figure, if PCs do not affect the detection and identification then there should be no change in the outputs. Therefore, by looking at the changes for different amounts of PCs we can see how the PC counts are affecting the algorithms, Figures \ref{fig:tarNorm} through \ref{fig:tarveh}.

In Figure \ref{fig:tarNorm}(a), the ACE scores of the ROIs are broken into targets in blue and false detections in orange.  The spectral fit, seen in Figure \ref{fig:tarNorm}(b), and probability, shown in Figure \ref{fig:tarNorm}(d), are part of our identification process. These panels show the probability of that the ROI is the identified target and the spectral fit of that ROI to the identified target. Figure \ref{fig:tarNorm}(e) shows which material the ROI had the highest probability of being identified as when compared to our target library. The visual location of the target, Figure \ref{fig:tarNorm}(c), and a more detailed visual, \ref{fig:tarNorm}(f), are provided.  As before, each $\times$ is the center of the ROI for a particular number of PCs used and the red pixels are the targets found using the full set of PCs. 

In Figure \ref{fig:tarNorm} the target can be found with most PC levels tested, but the reliability and quality of the detection start to decrease when fewer PCs are used.  In Figure \ref{fig:tarNorm}(e) the material identification for this ROI jumps between 3 different classifications but this ROI is correctly identified as F8, with all levels of PCs. The designation after F8 represents different measurements of the same material.

In Figure \ref{fig:tarSP} the small target goes undetected when using less than 70 PCs.  In general, the analysis of the whole data set suggests using 40 PCs would be sufficient for detection; however, such a low PC would be problematic for this target or targets of similar size. 

Figure \ref{fig:tarSG} is an example of another a small target. 
 Due to the small size of the target, detection becomes unreliable when using lower levels of PCs. With a small number of PCs the identification is inaccurate, predicting a different type of material for the target.  In this case, the material identification improves when a large number of PCs are used.

With Figure \ref{fig:tarTanCam}, the target can be consistently detected with a lower number of PCs but the false detection rate climes with fewer PCs. In this particular case, there are two materials for identification on the target, a vehicle with a fabric net, which may be why this ROI identification seems sensitive to changes in the level of PCs used. 

Lastly, the target in Figure \ref{fig:tarveh} is another type of vehicle. In this case, the target is always identified as the correct type of object; however, its detection rate is not consistent until about 35 PCs. It is also notable how much the spectral fit fluctuates with different numbers of PCs.

\section{Conclusion}
\label{sec:page}

Using the full number of PCs is redundant and the process may be sped up without sacrificing detection by using a fewer number of PCs. In general, using about 40 PCs does not seem to cause the ACE algorithm to miss the ROIs. However, reducing  the number of PCs used does cause higher ACE scores for non-targets. 

Although the probabilities are consistent across the different numbers of PCs the spectral fit tends to not work as well when using fewer PCs. This may be due to the pixels used to detect the background since those pixels are also based on the ACE scores. It is possible that one could account for this by adjusting the thresholds that are used to determine ROIs by growing and masking pixels for spectral fit. This would reinforce the spectral fit which should improve the identification output. 

Finally, it is necessary when deciding how many PCs to use to determine if target detection is appropriate or if further identification is important.  This may drastically impact the number of PCs required and will need to be balanced with the calculation speed, especially in a situation where real-time detection is needed.     

\section{Acknowledgments}

Thanks to Geospatial Technology Associates for access to the raw data used in the paper.


\bibliography{refs}
\bibliographystyle{IEEEbib}

\end{document}